\documentclass{physeauth}
\usepackage{graphicx}
\usepackage{amsmath}
\usepackage{amssymb}
\usepackage{epsfig}
\usepackage{bm}
\usepackage{latexsym}

\newcommand{\hide}[1]{}
\newcommand{\tbox}[1]{\mbox{\tiny #1}}

\newcommand{\trc}{\mbox{trace}}
\newcommand{\intt}{\int\!\!\!\!\int }

\newcommand{\eexp}{\mbox{e}^}

\newcommand{\ttimes} {\mbox{\tiny \ $^{\times}$ \ }}

\newcommand{\im}{\mbox{Im}}

\newcommand{\vect}[1]{\overrightarrow{#1}}

\newcommand{\Cn}[1]{\begin{center} #1 \end{center}} 
\newcommand{\be}{\begin{eqnarray}}
\newcommand{\ee}{\end{eqnarray}} 

\begin{document}


\begin{frontmatter}

\title{Quantum pumping and dissipation in closed systems}

\author{Doron Cohen}

\address{Department of Physics, Ben-Gurion University, 
Beer-Sheva 84105, Israel}


\begin{abstract}
Current can be pumped through a closed system
by changing parameters (or fields) in time.
Linear response theory (the Kubo formula)
allows to analyze both the charge transport
and the associated dissipation effect.
We make a distinction between adiabatic
and non-adiabatic regimes, and explain the
subtle limit of an infinite system.
As an example we discuss the following question:
What is the amount of charge which is pushed
by a moving scatterer?
In the low frequency (DC) limit we can
write $dQ = -G dX$, where $dX$ is the displacement
of the scatterer. Thus the issue is to calculate
the generalized conductance $G$.
\end{abstract}


\begin{keyword}

mesoscopics \sep quantum chaos \sep 
linear response \sep quantum pumping 

\thanks{Lecture notes for the {\em Physica~E} proceedings of the conference
"Frontiers of Quantum and Mesoscopic Thermodynamics" [Prague, July 2004].}


\PACS 03.65.-w 
\sep 73.23.-b 
\sep 05.45.Mt
\sep 03.65.Vf

\end{keyword}
 
\end{frontmatter}
 
 
\section{Introduction}
 
The analogy between electric current and the flow of water
is in fact older than the discovery of the electrons.
There are essentially two ways to move "water" (charge) between
two ``pools" (reservoirs): One possibility is to exploit
potential difference between the two reservoirs so as to
make the ``water" flow through a ``pipe" (wire). The other
possibility is to operate a device (pump) at some location along
the pipe (the ``scattering region"). This possibility of
moving charge without creating a potential difference is
called pumping. This description assumes ``open" geometry as
in Fig.1c. But what about a ``closed" system as in Fig.1b?
If we operate the same pump, do we get the same
circulating current as in the ``open" geometry?

\begin{figure}[h]
\Cn{\epsfig{figure=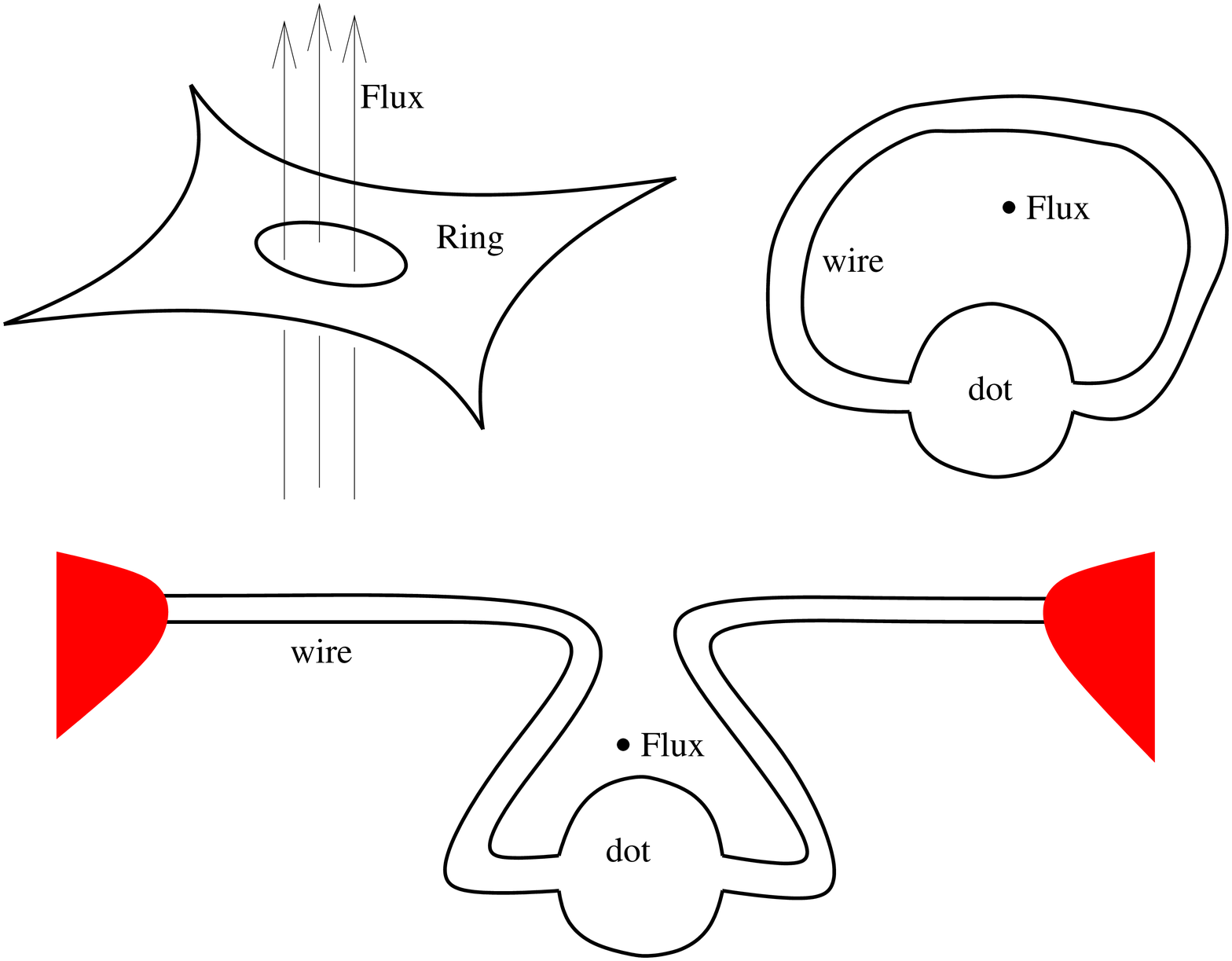,width=0.86\hsize}}
\caption{   
(a) Upper left: A chaotic ring that has 
the shape of a Sinai billiard, with Aharonov-Bohm flux. 
(b) Upper right: The dot-wire geometry with the same 
topology as in the case of the Sinai billiard. 
(c) Lower: The wire is cut into two leads that are attached 
to reservoirs. The latter is what we call ``open geometry".} 
\end{figure}

The analysis of ``quantum pumping" in closed systems
should take into account several issues that go beyond
the water analogy:
{\bf (i)} Kirchhoff law is not satisfied in the mesoscopic
reality because charge can accumulate;
{\bf (ii)} There are quantized energy levels,
consequently one has to distinguish between
adiabatic and non-adiabatic dynamics;
{\bf (iii)} Interference is important,
implying that the result of the calculation
is of statistical nature (universal conductance fluctuations).
On top we may have to take into account the effect
of having an external environment (decoherence).

Quantum pumping is a special issue in the study 
of ``driven systems". We are going to emphasize 
the significance of ``quantum chaos" in the analysis. 
This in fact provides the foundations for
linear response theory (LRT) 
\cite{landau,dsp,wilk,crs,frc,pmc}. 
We shall explain how to apply 
the Kubo formalism in order to analyze the dynamics 
in the low frequency (DC) regime. Within the Kubo 
formalism the problem boils down to the calculation 
of the generalized (DC) conductance matrix.

To avoid miss-understanding we emphasize that 
the dynamics in the low frequency (DC) regime 
is in general non-adiabatic: 
The DC conductance has both a dissipative 
and a non-dissipative parts. In the adiabatic 
limit (extremely small rate of driving) 
the dissipative part vanishes, while the 
non-dissipative part reduces to ``adiabatic transport" 
(also called ``geometric magnetism") 
\cite{berry,thouless,AvronNet,BeRo}. 
The ``adiabatic regime", 
where the dissipative effect can be ignored, 
is in fact a tiny sub-domain 
of the relatively vast ``DC regime".

The dot-wire geometry of Fig.1b is of particular
interest. We are going to discuss the special limit
of taking the length of the wire ($L$) to be infinite.
In this limit the adiabatic regime vanishes,
but still we are left with a vast "DC regime" where
the pumping is described by a "DC conductance".
In this limit we get results \cite{pmo} 
that are in agreement
with the well known analysis of quantum pumping 
\cite{BPT,brouwer} in an open geometry (Fig.1c).

\begin{figure}[b]
\Cn{\epsfig{figure=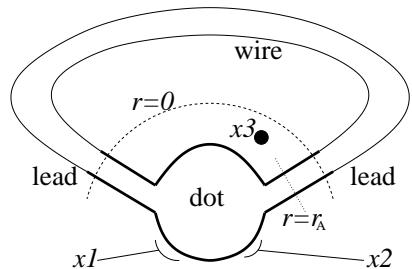,width=0.7\hsize}}
\caption{   
Detailed illustration of the dot-wire system.
The dot potential is controlled
by gate voltages $X_1$ and $X_2$.
The flux through the loop is $X_3{=}\Phi$.
The scattering region ($r{<}0$)
is represented by an $S$~matrix. 
Later we assume that the length ($L$) 
of the wire is very large.}

\end{figure}

\section{Driven systems}

Consider a Fermi sea of non interacting ``spinless" electrons. 
The electrons are bounded by some potential. To be specific 
we assume a ring topology as in Fig.1a. Of particular 
interest is the dot-wire geometry of Fig.1b, or its more 
elaborated version Fig.2. It has the same 
topology but we can distinguish between a ``wire region" and 
a ``dot region" (or ``scattering region"). 
In particular we can consider a dot-wire system such 
that the length of the wire is very very long. 
If we cut the wire in the middle, and attach each lead 
to a reservoir, then we get the open geometry of Fig.1c.

We assume that we have some control over 
the potential that holds the electrons. 
Specifically, and without loss of generality, 
we assume that there are control parameters $X_1$ and $X_2$ 
that represent e.g. some gate voltages (see Fig.2) 
with which we can control the potential 
in the scattering region.
Namely, with these parameters we can change the 
dot potential floor, or the height of some 
barrier, or the location of a ``wall" element, 
or the position of a scatterer inside the dot.
We call $X_1$ and $X_2$ shape parameters.

We also assume that it is possible to have 
an Aharonov-Bohm flux $X_3$ through the ring. 
Thus our notations are:
\be
& & X_1, X_2 \ = \ \mbox{shape parameters} \\
& & X_3 \ = \ \Phi \ = \ (\hbar/e)\phi \ = \ \mbox{magnetic flux}
\ee
and the motion of each electron is described 
by a one particle Hamiltonian 
\be
\mathcal{H} \ = \ \mathcal{H}(\bm{r},\bm{p};\ X_1(t),X_2(t),X_3(t))
\ee

To drive a system means to change 
some parameters (fields) in time.
No driving means that $X_1$ and $X_2$ are 
kept constant, and also let us assume for simplicity 
that there is no magnetic field and that $X_3=0$. 
In the absence of driving we assume 
that the motion of the electrons inside 
the system is classically chaotic. 
For example this is the case with the 
so-called Sinai billiard of Fig.1a. 
In such circumstances the energy of the system 
is a constant of the motion, and the net circulating 
current is zero due to ergodicity.

The simplest way to create a current $\mathcal{I}$ 
in an open system (Fig.1c) is to impose bias 
by having a different chemical potential in each reservoir.
Another possibility is to create 
an electro-motive-force (EMF) in the dot region. 
In linear response theory it can be proved 
that it does not matter what is the assumed 
distribution of the voltage along the ``resistor". 
The EMF is by Faraday law $-\dot{\Phi}$. 
Assuming DC driving (constant EMF), 
and the applicability of LRT, we get the ``Ohm law" 
$\mathcal{I} = \bm{G}^{33} \times (-\dot{\Phi})$ 
and hence the transported charge is 
$dQ = - \bm{G}^{33} \ dX_3$. 
We call $\bm{G}^{33}$ the Ohmic (DC) conductance.
If we have a low frequency AC driving rather 
than a DC driving, still the impedance (AC conductance) 
is expected to be well approximated by the DC conductance 
within a frequency range that we call the DC regime.

Yet another possibility is to induce current by 
changing shape parameter in time, 
while keeping either the bias or $X_3$ equal to zero.
Say that we change $X_1$, then in complete 
analogy with Ohm law we can write 
$dQ = - \bm{G}^{31} \ dX_1$. More generally 
we can write 
\be
dQ \ = \ - \sum_j \bm{G}^{3j} \ dX_j
\ee
Obviously this type of formula makes sense 
only in the ``DC regime" where the current 
at each moment of time depends only on the 
rates $\dot{X}_j$.

\begin{figure}[b]
\Cn{
\epsfig{figure=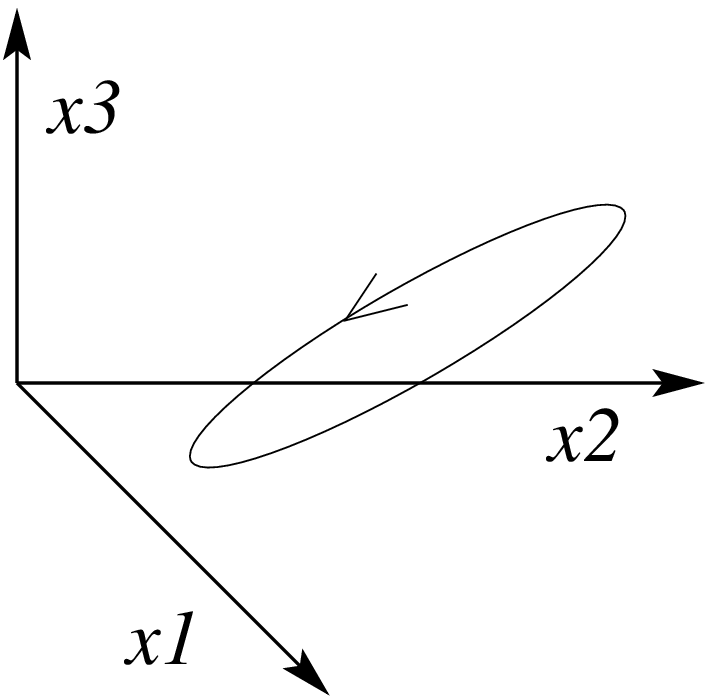,height=0.35\hsize} 
\ \ \ \ 
\epsfig{figure=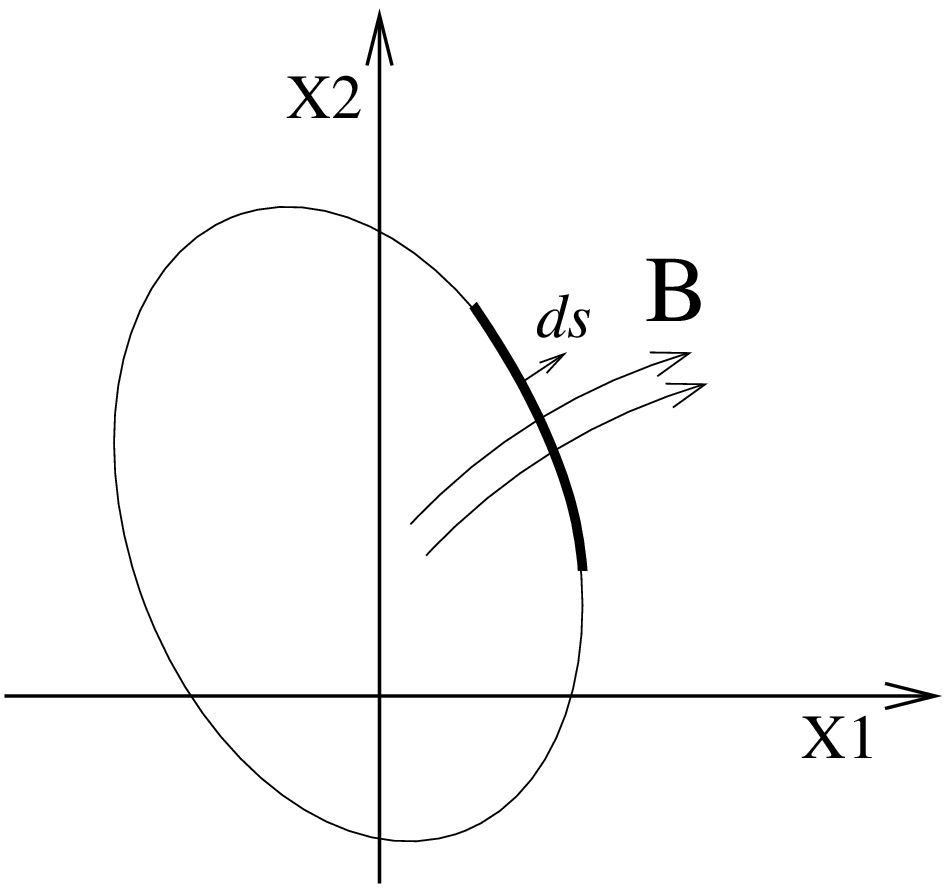,height=0.35\hsize}
}
\caption{   
(a) Left: A driving cycle in $X$ space. In order to have non-zero 
area enclosed we have to change (without loss of generality)  
two parameters. (b) Right: In particular we consider pumping cycle 
in the $X_3=0$ plane (no magnetic field). }
\end{figure}

\section{pumping cycles}

In practice the interest is a time periodic (AC) driving. 
This means that the driving cycle can be represented 
by a closed contour at the $(X_1,X_2,X_3)$ space
as in Fig.3a. In fact we assume that the contour is 
lying in the $(X_1,X_2)$ plan as in Fig.3b.
We ask what is the amount of charge which is transported 
via a section of the ring per cycle. 
Assuming the applicability of LRT we get in the DC regime 
\be \label{e5}
Q \ = \ \oint \mathcal{I} dt \ = \ \oint \bm{G} \cdot dX
\ee 
where $X=(X_1,X_2,X_3)$ and 
$\bm{G} = (\bm{G}^{31},\bm{G}^{32},\bm{G}^{33})$. 
Later we shall define a more general object $\bm{G}^{kj}$
with $k,j=1,2,3$ that we call {\em generalized conductance matrix}. 
In the above formula only the $k=3$ 
row enters into the calculation.

Getting $Q\ne 0$ means that the 
current has a non-zero DC component. 
So we can define ``pumping" as 
getting DC current form AC driving. 
From the above it is clear that 
within the DC regime we have to vary  
at least two parameters to 
achieve a non-zero result. 
In a closed (in contrast to open) 
system this conclusion remains 
valid also outside of the DC regime, 
due to time reversal symmetry.
In order to get DC current from one parameter 
AC driving, in a closed system, 
it is essential to have a non-linear response.
{\em Ratchets} are non-linear devices 
that use ``mixed" \cite{ratchH}   
or ``damped" \cite{ratchD} dynamics 
in order to pump with only one parameter.
We are {\em not} discussing such devices below.


\section{What is the problem?}

Most of the studies of quantum pumping were (so far)
about open systems. Inspired by Landauer who pointed out 
that $\bm{G}^{33}$ is essentially the transmission of the device,
B{\"u}ttiker, Pretre and Thomas (BPT) have
developed a formula that allows the calculation of $\bm{G}^{3j}$ 
using the $S$ matrix of the scattering region \cite{BPT,brouwer}.
It turns out that the non-trivial extension of this approach
to closed systems involves quite restrictive assumptions \cite{MoBu}.
Thus the case of pumping in closed systems has been left un-explored,
except to some past works on adiabatic transport \cite{AvronNet,BeRo}.
Yet another approach to quantum pumping is to use
the powerful {\em Kubo~formalism} \cite{pmc,pmo,pmt}.

The Kubo formula, which we discuss later, 
gives a way to calculate the 
generalized conductance matrix $\bm{G}^{kj}$. 
It is a well know formula \cite{landau}, 
so one can ask: what is the issue here? 
The answer is that both the validity conditions, 
and also the way to use the Kubo formula, 
are in fact open problems in physics.

The Van Kampen controversy regarding the 
validity of the Kubo formula in the classical 
framework is well known, and by now has 
been resolved. For a systematic classical derivation 
of the Kubo formula with all the validity 
conditions see Ref.\cite{frc} and references therein. 
The assumption of chaos is essential 
in the classical derivation. 
If this assumption is not satisfied 
(as in the trivial case of a driven 1D ring) 
then the Kubo formula becomes non-applicable.

What about the Quantum Mechanical derivation? 
The problem has been raised in Ref.\cite{wilk} 
but has been answered only later 
in Refs.\cite{crs,frc} and follow up works.
It is important to realize that the quantum 
mechanical derivation of the Kubo formula 
requires perturbation theory to infinite order, 
not just 1st order perturbation theory. 
We shall discuss later the non-trivial 
self consistency condition of the quantum mechanical 
derivation.

We note that the standard textbook derivation 
of the Kubo formula assumes that the 
energy spectrum is essentially a continuum. 
A common practice is to assume some weak 
coupling to some external bath \cite{imryK}. 
However, this procedure avoids the question 
at stake, and in fact fails to take into 
consideration important ingredients that 
have to do with {\em quantum chaos physics}.
In this lecture the primary interest 
is in the physics of a closed {\em isolated} system.  
Only in a later stage we look for the effects 
that are associated with having a weak coupling 
to an external bath.

Why do we say that it is not clear how 
to use the Kubo formula? We are going to explain 
that the quantum mechanical derivation of the 
Kubo formula introduces an energy scale 
that we call $\Gamma$. It plays an analogous 
role to the level broadening parameter 
which is introduced in case of a coupling to a bath. 
Our $\Gamma$ depends on the rate $\dot{X}$ 
of the driving in a non-trivial way.
One may say that $\Gamma$ in case of an isolated 
system is due to the non-adiabaticity of the driving.
Our $\Gamma$ affects both the dissipative 
and the non-dissipative (geometric) part 
of the response. Without a theory for 
$\Gamma$ the quantum mechanical Kubo formula 
is ill defined.

\section{Generalized forces and currents}

Given a Hamiltonian we define generalized forces 
in the conventional way:
\be
\mathcal{F}^k \ = \ -\frac{\partial \mathcal{H}}{\partial X_k}
\ee
one obvious reasoning that motivates this definition 
follows from writing the following (exact) expression for 
the change in the energy $E=\langle \mathcal{H} \rangle$ 
of the system:
\be
E_{\tbox{final}}-E_{\tbox{initial}} \ = \ 
- \int \langle \mathcal{F}(t) \rangle \cdot dX
\ee
In particular we note that $\mathcal{F}^3$ should be 
identified as the current $\mathcal{I}$. 
This identification can be explained as follows: 
If we make a change $d\Phi$ of the flux during a time $dt$, 
then the EMF is $-d\Phi/dt$, leading to a current $\mathcal{I}$. 
The energy increase is the EMF times the charge, 
namely $dE=(-d\Phi/dt)\times(\mathcal{I}dt)=-\mathcal{I}d\Phi$.
Hence $\mathcal{I}$ is conjugate to $\Phi$.

As an example we consider \cite{pmt}  
a network model \cite{kottos}.  
See the illustration of Fig.4d. 
The Hamiltonian is 
\be
\mathcal{H} \ \ = \ \ \mbox{\small network} 
\ \ + \ \ X_2 \ \delta(x-X_1)
\ee
We assume control over the position $X_1$ 
of the delta scatterer, 
and also over the ``height" $X_2$ 
of the scatterer. By the definition we get:
\be
\mathcal{F}^1 \ &=& \  X_2 \delta'(x-X_1)
\\ 
\mathcal{F}^2 \ &=& \  -\delta(x-X_1)
\ee
Note that $\mathcal{F}^1$ is the ordinary Newtonian force 
which is associated with translations. Its operation on 
the wavefunction can be realized by the differential operator  
\be
\mathcal{F}^1 \ \ \mapsto \ \ -X_2 
\left( \overrightarrow{\partial} + \overleftarrow{\partial} 
- \frac{2\mathsf{m}}{\hbar^2} X_2 \right)_{x=X_1+0}
\ee
where we have used the matching condition across the delta 
function and $\mathsf{m}$ is the mass of the particle.

What about the current operator? For its definition we have 
to introduce a vector potential $\mathcal{A}(x) = \Phi a(x)$  
into the Hamiltonian such that 
\be
\oint \vect{\mathcal{A}} \cdot \vect{dr} \ = \ \Phi
\ee 
Thus we have to specify $a(x)$, which describes how 
the vector potential varies along the loop.   
This is not merely a gauge freedom because the 
electric field $-\dot{\Phi} a(x)$ is a measurable 
quantity. Moreover, a different $a(x)$ implies 
a different current operator. In particular we can 
choose $a(x)$ to be a delta function across 
a section $x=x_0$. Then we get:
\be
\mathcal{I} \ \ = \ \ \frac{e}{2\mathsf{m}} 
\left( \delta(x-x_0)p + p\delta(x-x_0) \right)
\ee
Note that the operation of this operator 
can be realized by the differential operator
\be
\mathcal{I} \ \ \mapsto \ \ -i \frac{e\hbar}{2\mathsf{m}} 
\Big(\overrightarrow{\partial}-\overleftarrow{\partial}\Big)_{x=x_0} 
\ee
A few words are in order regarding the continuity of the charge flow.
It should be clear that in any moment the current through 
different sections of a wire does not have to be the same, 
because charge can accumulate. Kirchhoff law is not satisfied. 
For example if we block the left entrance to the dot in Fig.2, 
and raise the dot potential, then current is pushed out of 
the right lead, while the current in the blocked side is zero. 
Still if we make a full pumping cycle, such that the charge 
comes back to its original distribution at the end of each cycle, 
then the result for $Q$ should be independent of the section 
through which the current is measured.

\begin{figure}[b]
\epsfig{figure=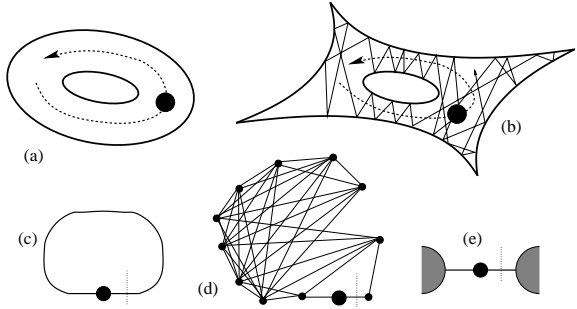,width=\hsize}
\caption{   
A scatterer (represented by a black circle) is 
translated through a system that has a Fermi occupation 
of spinless non-interacting electrons. 
In (a) the system is a simple ring. 
In (b) it is a chaotic ring (Sinai billiard). 
In (c) and in (d) we have network systems that 
are of the same type of (a) and (b) respectively. 
In the network, the scatterer (``piston") 
is a delta function (represented as a big circle) located at $x=X_1$. 
The current is measured through $x=x_0$ (dotted vertical line).
In (e) we have an open geometry with left and right leads that 
are attached to reservoirs that have the same chemical potential.} 
\end{figure}

\section{Linear response theory}

Assume that $X(t)=X^{(0)} + \delta X(t)$, 
and look for a quasi-stationary solution.
To have linear response means that the generalized 
forces are related to the driving as follows: 
\be \label{e15}
\langle \mathcal{F}(t) \rangle \ = \ 
\langle \mathcal{F} \rangle_0  \ + \ 
\int_{-\infty}^{\infty} 
\bm{\alpha}(t-t') \cdot \delta X(t') \ dt' 
\ee 
where $\langle ... \rangle_0$ denote the expectation 
value with respect to the unperturbed $X(t)=X^{(0)}$ 
stationary state. From now on we disregard the zero 
order term (the ``conservative force"), and focus 
on the linear term. 
The generalized susceptibility $\chi^{kj}(\omega)$ 
is the Fourier transform of the (causal) response 
kernel $\alpha^{kj}(\tau)$, while the generalized 
conductance matrix is defined as 
\be
\bm{G}^{kj} \ = \ \left. 
\frac{\mbox{Im}[\chi^{kj}(\omega)]}{\omega} 
\ \right|_{\omega\sim0} \ = \ \bm{\eta}^{kj} + \bm{B}^{kj}
\ee
The last equality defines the symmetric 
and the anti-symmetric matrices $\bm{\eta}^{kj}$ and $\bm{B}^{kj}$. 
Thus in the DC limit Eq.(\ref{e15}) reduces to a generalized Ohm law: 
\be
\langle \mathcal{F}^k \rangle \ \ = \ \
-\sum_{j} \bm{G}^{kj} \ \dot{X}_j
\ee
which can be written in fancy notations as 
\be
\langle F \rangle \ \ = \ \ -\bm{G}\cdot \dot{X} \ \ = \ \ 
-\bm{\eta} \cdot \dot{X} \ - \ \bm{B}\wedge \dot{X}
\ee
Note that the rate of dissipation is 
\be
\dot{\mathcal{W}} \ \ = \ \ -\langle F \rangle \cdot \dot{X} \ \ = \ \
\sum_{kj} \bm{\eta}^{kj} \ \dot{X}_k \ \dot{X}_j
\ee

We would like to focus not on the dissipation issue,  
but rather on the transport issue. From Eq.(\ref{e5}) we get 
\be
Q \ \ = \ \ 
\Big[ \ -\oint \bm{\eta} \cdot dX \ \ - \oint \bm{B} \wedge dX \ \ \Big]_{k=3}
\ee
From now on we consider a planar $(X_1,X_2)$ pumping cycle, 
and assume that there is no magnetic field. 
Then it follows from time reversal symmetry [Onsager] that 
$\bm{\eta}^{31} = \bm{\eta}^{32} = 0$, and consequently 
\be \label{e21}
Q \ = \ -\oint \vect{\bm{B}} \cdot \ \vect{ds} 
\ee
where $\vect{\bm{B}}=(\bm{B}^{23},\bm{B}^{31},\bm{B}^{12})$, 
with $\bm{B}^{12} = 0$, and $\vect{ds}=(dX_2,-dX_1,0)$ 
is a normal vector in the pumping plane as in Fig.3b.

\newpage

The various objects that have been defined  
in this section are summarized by the following diagram:

\ \\
{
\setlength{\unitlength}{2000sp}
\begin{picture}(4725,5767)(751,-7112)
\put(1501,-1861){\vector( 0,-1){375}}
\put(1501,-3061){\vector(-2,-3){242.308}}
\put(1801,-3061){\vector( 4,-1){1641.177}}
\put(4801,-4261){\vector( 0,-1){525}}
\put(4951,-5611){\vector( 1,-1){375}}
\put(4651,-5611){\vector(-1,-1){375}}
\put(1126,-1561){$\alpha^{kj}(t-t')$}
\put(1201,-2761){$\chi^{kj}(\omega)$}
\put(751,-3886){$\mbox{Re}[\chi^{kj}(\omega)]$}
\put(3301,-3886){$(1/\omega) \ttimes \mbox{Im}[\chi^{kj}(\omega)]$}
\put(3676,-6511){$\bm{\eta}^{kj}$}
\put(5326,-6511){$\bm{B}^{kj}$}
\put(5476,-7036){(non-dissipative)}
\put(4576,-5386){$\bm{G}^{kj}$}
\put(3001,-7036){(dissipative)}
\end{picture}
}
\ \\

\section{The Kubo formula}

The Kubo formula for the response kernel is 
\be
\alpha^{kj}(\tau) \ = \ \Theta(\tau) \times 
\frac{i}{\hbar} \langle [\mathcal{F}^k(\tau),\mathcal{F}^j(0)]\rangle_0
\ee
where the expression on the right hand side 
assumes a zero order $X=X^{0}$ stationary state 
(the so called ``interaction picture"), 
and $\Theta(\tau)$ is the step function.

Using the definitions of the previous section, 
and assuming a Fermi sea of non-interacting fermions  
with occupation function $f(E)$,  
we get the following expressions:
\be \nonumber 
\bm{\eta}^{kj} &=& 
-\pi\hbar\sum_{n,m}
\frac{f(E_n){-}f(E_m)}{E_n{-}E_m}
\mathcal{F}^k_{nm}\mathcal{F}^j_{mn}
\ \delta_{\Gamma}(E_m{-}E_n)
\\ \label{e23}
\bm{B}^{kj} &=& 
2\hbar \sum_n f(E_n)
\sum_{m(\ne n)}
\frac{\im\left[\mathcal{F}^k_{nm}\mathcal{F}^j_{mn}\right]}
{(E_m{-}E_n)^2+(\Gamma/2)^2}
\ee
We have incorporated in these expression 
a broadening parameter $\Gamma$ which is absent 
in the ``literal" Kubo formula. If we set 
$\Gamma=0$ we get no dissipation 
($\bm{\eta}=0$). We also see that $\Gamma$ affects 
the non-dissipative part of the response. 
Thus we see that without having a theory 
for $\Gamma$ the Kubo formula is an ill defined expression.

\section{Adiabatic transport (Geometric magnetism)}

The ``literal" Kubo formula (i.e. with $\Gamma=0$) 
has been considered in Refs.(\cite{AvronNet,BeRo}). 
In this limit we have no dissipation ($\bm{\eta}=0$). 
But we may still have a non-vanishing $\bm{B}$.
By Eq.(\ref{e23}) the total $\bm{B}$ is a sum 
over the occupied levels. The contribution of 
a given occupied level $n$ is:
\be
\bm{B}^{kj}_n \ \ = \ \
2\hbar 
\sum_{m(\ne n)}
\frac{\im\left[
\mathcal{F}^k_{nm}\mathcal{F}^j_{mn}\right]}
{(E_m-E_n)^2+(\Gamma/2)^2}
\ee
with $\Gamma=0$. This is identified as the 
geometric magnetism of Ref.\cite{BeRo}.

We can get some intuition for $\vect{\bm{B}}$  
from the theory of adiabatic processes. 
The Berry phase is given as a line integral 
$ (1/\hbar)\oint \vect{\bm{A}} \cdot dX $ over ``vector potential" 
in $X$ space. By stokes law it can be converted 
to an integral  $(1/\hbar)\intt \vect{\bm{B}} \cdot dS$   
over a surface that is bounded by the driving cycle.  
The $\vect{\bm{B}}$ field is divergence-less, but 
it may have singularities at $X$ points where 
the level $n$ has a degeneracy with a nearby level.  
We can regard these points as the location of 
magnetic charges.  The result of the surface integral 
should be independent of the choice of the surface modulo $2\pi$, 
else Berry phase would be ill defined. Therefore 
the net flux via a closed surface (which we can regard as 
formed of two Stokes surfaces) should be zero modulo $2\pi$.
Thus, if we have a charge within a closed 
surface it follows by Gauss law that it should 
be quantized in units of $(\hbar/2)$. These are the 
so called ``Dirac monopoles".  In our setting $X_3$ 
is the Aharonov-Bohm flux. Therefore we have 
vertical ``Dirac chains"
\be
\mbox{chain} \ = \ \left(X_1^{(0)}, \ \ X_2^{(0)}, \ \ 
\Phi^{(0)}+2\pi\frac{e}{\hbar}\times\mbox{\small integer}\right)
\ee
In the absence of any other magnetic field we have 
time-reversal symmetry for either integer or half integer flux.
It follows that there are two types of Dirac chains: 
those that have a monopole in the plane of the
pumping cycle, and those that have their monopoles 
half unit away from the pumping plane.

In the next section we shall see how these observations 
help to analyze the pumping process. We shall also illuminate 
the effect of having $\Gamma \ne 0$. 
Later we shall discuss the ``physics" behind $\Gamma$.

\section{Quantized pumping?}

The issue of quantized pumping is best illustrated 
by the popular two delta barrier model, 
which is illustrated in Fig.5.  
The ``dot region" $|Q|<a/2$ is described by the potential
\begin{eqnarray}
U(r;X_1,X_2) =
X_1\delta\left(x+\frac{a}{2}\right) +
X_2\delta\left(x-\frac{a}{2}\right)
\end{eqnarray}
The pumping cycle is described in Fig.5c.
In the 1st half of the cycle an electron 
is taken from the wire into the dot region 
via the left barrier,  while in the second 
half of the cycle an electron is transfered 
from the dot region to the wire 
via the right barrier. 
So it seems that one electron is pumped 
through the device per cycle. 
The question is whether it is exactly 
one electron ($Q=e$) or not?

\begin{figure}[b]
\epsfig{figure=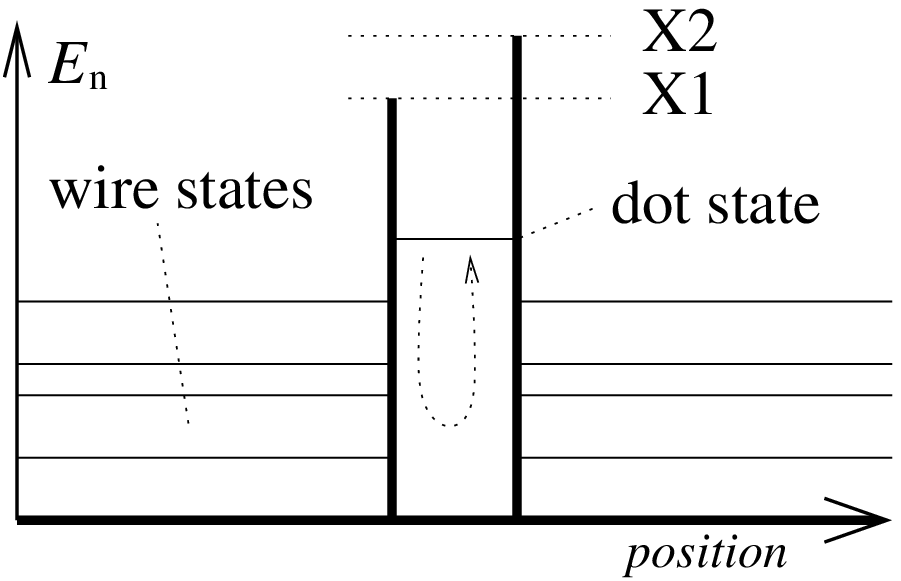,width=0.45\hsize}
\ \ \ \
\epsfig{figure=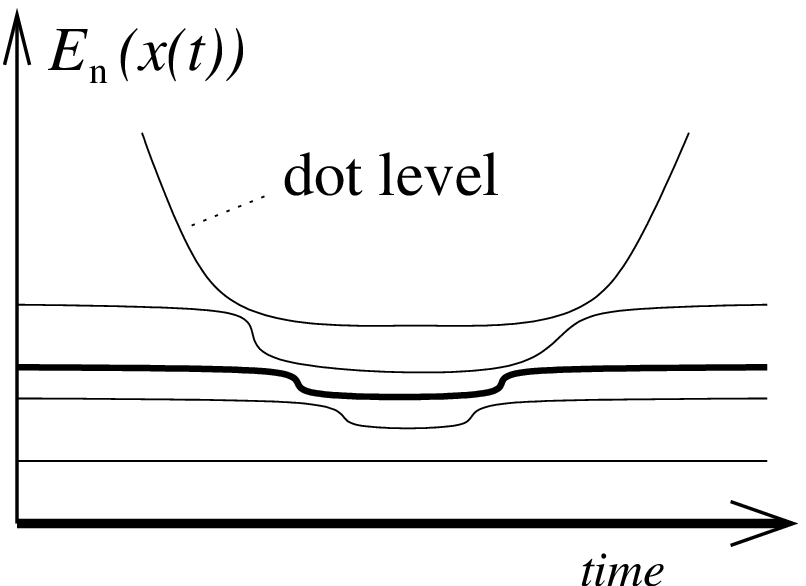,width=0.45\hsize} \\
\Cn{\epsfig{figure=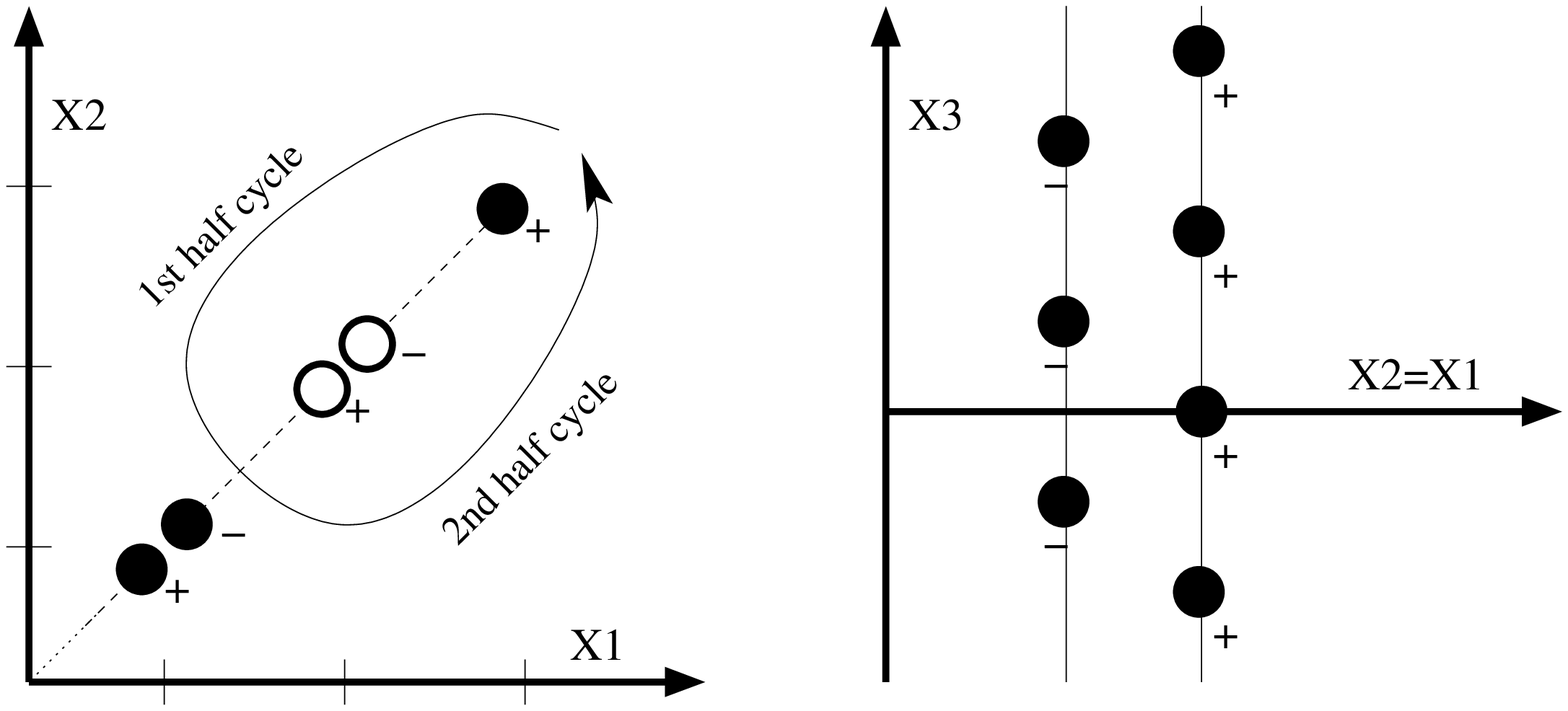,width=0.90\hsize}}
\caption{   
(a) Upper left: The energy levels of a ring 
with two barriers, at the beginning of the pumping cycle. 
It is assumed that the three lower levels are occupied. 
(b) Upper right: The adiabatic levels as a function 
of time during the pumping cycle. 
(c) Lower Left: The $(X_1,X_2)$ locations of 
the Dirac chains of the $3$ occupied levels. 
Filled (hollow) circles imply that there 
is (no) monopole in the pumping plane. 
Note that for sake of illustration overlapping 
chains are displaced from each other.
The pumping cycle encircles $2+1$ Dirac chains 
that are associated with the 3rd and 2nd levels respectively.   
(d) Lower right: The $2$ Dirac chains 
that are associated with the 3rd level.}
\end{figure}

In the case of an open geometry the answer is known 
\cite{SAA,barriers}.
Let us denote by $g_0$ the average transmission 
of the dot region for $X$ values along the pumping 
cycle. In the limit $g_0 \rightarrow 0$, which 
is a pump with no leakage, indeed one gets $Q=e$. 
Otherwise one gets $Q=(1-g)e$.

What about a {\em closed} (ring) geometry? 
Do we have a similar result?   
It has been argued \cite{SAA} 
that if the the pumping process is strictly adiabatic 
then we get exactly $Q=e$. We are going to explain 
below that this is in fact not correct:  
We can get either $Q<1$ or $Q>1$ or even $Q \gg 1$.

Recall that by Eq.(\ref{e21}) the pumped charge $Q$ 
equals the projected flux of the $\vect{\bm{B}}$ field 
through the pumping cycle (Fig.3b). 
If the charge of the monopoles 
were uniformly distributed along the chains, it would 
follow that $Q$ is exactly quantized. 
But this is not the case, 
and therefore $Q$ can be either 
smaller or larger than $1$ depending 
on the type of chain(s) being encircled.  
In particular, in case of a tight cycle 
around a monopole we get $Q\gg e$ which 
is somewhat counter-intuitive, while if 
the monopole is off-plane $Q < e$.

What is the effect of $\Gamma$ on this result? 
It is quite clear that $\Gamma$ diminishes 
the contribution of the singular term. 
Consequently it makes $Q$ less than one. 
This gives us a hint that the introduction of 
$\Gamma$ might lead to a result which is 
in agreement with that obtained for an open geometry. 
We shall discuss this issue in the next sections.

\section{The Kubo Formula and ``quantum chaos"}

We turn now to discuss $\Gamma$. Any generic 
quantum chaos system is characterized by some 
short correlation time $\tau_{cl}$, 
by some mean level spacing $\Delta$, 
and by a semiclassical energy scale 
that we denote as $\Delta_b$. Namely:
\be 
\Delta \ & \propto& \ \hbar^{d}/\mbox{\small volume} \ = 
\ \mbox{\small mean level spacing} \\
\Delta_b \ & \sim & \ \hbar/\tau_{\tbox{cl}} \ = 
\ \mbox{\small bandwidth}
\ee
The term bandwidth requires clarification. 
If we change a parameter $X$ in the Hamiltonian $\mathcal{H}$, 
then the perturbation matrix $\mathcal{F}_{nm}$ 
has non-vanishing matrix elements within 
a band $|E_n-E_m|<\Delta_b$. These matrix elements 
are characterized by some root-mean-square magnitude $\sigma$, 
while outside of the band the matrix elements are very small.

If the system is driven slowly in a rate $\dot{X}$
then levels are mixed non-perturbatively. 
Using a quite subtle reasoning \cite{crs,frc,pmc,dsp} 
the relevant energy range for the non-perturbative 
mixing of levels is found to be 
\be \label{e29}
\Gamma \ \ = \ \
\left(\frac{\hbar\sigma}{\Delta^2}|\dot{X}|\right)^{2/3} \!\times \Delta 
\ \ \ \ \propto \ \ \ \ 
\left(L \frac{}{} |\dot{X}|\right)^{2/3} \frac{1}{L} 
\ee
The latter equality assumes dot-wire geometry as in 
Fig.1b, where $L$ is the length of the wire.
Now we can distinguish between three $\dot{X}$ regimes: 
\be
\Gamma \ll \Delta & \ &  \ \ \ \mbox{adiabatic regime} \\
\Delta <  \Gamma  < \Delta_b & \ & \ \ \ \mbox{non-adiabatic regime} \\
\mbox{\em otherwise} & \ & \ \ \ \mbox{non-perturbative regime}
\ee
In the adiabatic regime levels are not mixed by the driving, 
which means that the system (so to say) follows the same level 
all the time. In the perturbative regime there is 
a non-perturbative mixing on small energy scales, but 
on the large scale we have Fermi-Golden-Rule (FGR) transitions.
If the self consistency condition ($\Gamma \ll \Delta_b$) 
breaks down, then the FGR picture becomes non-applicable, 
and consequently $\Gamma$ becomes a meaningless parameter.

In the non-perturbative regime we expect semiclassical methods 
to be effective, provided the system has a classical limit 
(which is not the case with random matrix models \cite{rsp}).
In general one can argue that in the limit of infinite volume 
(or small $\hbar$) perturbation theory always breaks down, 
leading to a semiclassical behavior. But in the dot-wire geometry 
this is not the case if we take the limit $L\rightarrow\infty$, 
keeping the width of the wire fixed. With such limiting procedure 
Eq.(\ref{e29}) implies that the self-consistency condition 
$\Gamma \ll \Delta_b$ is better and better satisfied! 
This means that the Kubo formula can be trusted. 
Furthermore, with the same limiting procedure 
the $L\rightarrow\infty$ is a {\em non-adiabatic} limit 
because the adiabaticity condition $\Gamma \ll \Delta$ breaks down.

\section{Kubo formula using an FD relation}

The Fluctuation-dissipation (FD) relation 
allows us to calculate the conductance $\bm{G}^{kj}$ 
from the correlation function $C^{kj}(\tau)$ 
of the generalized forces.  
In what follows we use the notations:
\be
K^{kj}(\tau) \ &=& \ 
\frac{i}{\hbar} \langle [\mathcal{F}^k(\tau),\mathcal{F}^j(0)]\rangle_0 
\\
C^{kj}(\tau) \ &=& \
\frac{1}{2} 
\left( \langle \mathcal{F}^k(\tau)\mathcal{F}^j(0) \rangle_0 + cc\right) 
\ee
Their Fourier transforms are denoted 
$\tilde{K}^{kj}(\omega)$ and $\tilde{C}^{kj}(\omega)$. 
The expectation value above assumes a zero order 
stationary preparation. 
We shall use subscript $|_F$ to indicate many-body Fermi occupation. 
We shall use  the subscript $|_T$ or the subscript $|_E$ 
to denote one-particle canonical or microcanonical preparation. 
At high temperatures the Boltzmann approximation applies 
and we can use the exact relation 
$f(E_n){-}f(E_m) = \tanh((E_n{-}E_m)/(2T)) \times (f(E_n){+}f(E_m))$
so as to get
\be
\tilde{K}^{kj}_F(\omega) = i\omega \times
\frac{2}{\hbar\omega}\tanh\left(\frac{\hbar\omega}{2T}\right)  \  C^{kj}_T(\omega)
\ee
At low temperatures we can use the approximation  
$f(E){-}f(E') \approx -\frac{1}{2}[\delta_T(E{-}E_F)+\delta_T(E'{-}E_F)] \times (E{-}E')$
with $\delta_T(E{-}E_F)=-f'(E)$ so as to get 
\be
\tilde{K}^{kj}_F(\omega) & \approx & 
i \omega \times g(E) \ \tilde{C}^{kj}_{E_F}(\omega)
\ee
The application of this approximation 
is ``legal" if we assume temperature $T\gg \Delta_b$. 
This is a very ``bad" condition because for (e.g.) ballistic 
dot $\Delta_b$ is the relatively large 
Thouless energy. However, we can regard the 
large $T$ result as an  $E_F$~averaged  
zero temperature calculation. Then it can 
be argued that for a quantum chaos system with 
a generic bandprofile the average is in fact 
the ``representative" result (see discussion 
of ``universal conductance fluctuation" in later sections).

Substituting the Kubo formula 
$\alpha^{kj}(\tau) = \Theta(\tau) \ K^{kj}(\tau)$
in the definition of $\bm{G}^{kj}$, 
and using the latter relation 
between $K^{kj}(\tau)$ and  $C^{kj}(\tau)$
we get after some straightforward algebra   
the following expression for the conductance:
\be \label{e37}
\bm{G}^{kj} =
\int_0^{\infty} K_F^{kj}(\tau)\tau d\tau
\ \approx \ 
\mathsf{g}(E_F) \int_0^{\infty}C_{E_F}^{kj}(\tau)d\tau 
\ee
where $\mathsf{g}(E_F)$ is the density of the one-particle states.
If we want to incorporate $\Gamma$ the recipe is simply:
\be \label{e38}
C(\tau) \ \ \mapsto \ \ 
C(\tau) \ \eexp{-\frac{1}{2}(\Gamma/\hbar) |\tau|} 
\ee

The expression of $\bm{G}^{kj}$ using $C^{kj}(\tau)$
is a generalized FD relation. It reduces to the 
standard FD relation if we consider the dissipative part: 
\be
\bm{\eta}^{kj} \ = \ \frac{1}{2} \mathsf{g}(E_F) \tilde{C}_{E_F}^{kj}(\omega \sim 0) 
\ee
whereas the non-dissipative part requires integration 
over all the frequencies (see next section).

\section{Kubo via Green functions or $S$ matrix}

Now we would like to express $\bm{G}^{kj}$ using Green functions, 
and eventually we would like to express it using the $S$ matrix 
of the scattering region. The first step is to rewrite  
the FD relation as follows:
\be
\bm{G}^{kj} \ \ &=& \ \ 
\hbar \mathsf{g}(E_F) \int_{-\infty}^{\infty} 
\frac{-i\tilde{C}_{E_F}^{kj}(\omega)}{\hbar\omega-i(\Gamma/2)} 
\ \frac{d\omega}{2\pi}
\ee
The second step is to write
\be
C^{kj}_E(\omega) = \frac{\hbar}{2\mathsf{g}(E)}
\left[ C^{kj}(E{+}\hbar\omega,E) + C^{jk}(E{-}\hbar\omega,E) \right]
\ee
where
\be
C^{kj}(E',E) &=& 
2\pi \sum_{nm} 
\mathcal{F}^k_{nm} \delta(E'-E_m) 
\mathcal{F}^j_{mn} \delta(E-E_n) 
\\
&=&
\frac{2}{\pi} 
\ \trc\left[ 
\mathcal{F}^k  \ \im[\mathsf{G}(E')] \ \mathcal{F}^j \ \im[\mathsf{G}(E)] 
\right]
\ee
We use the standard notations 
$\mathsf{G}(z)=1/(z-\mathcal{H})$,  
and $\mathsf{G}^{\pm}(E)=G(E{\pm}i0)$, 
and $\im[\mathsf{G}]=-i(\mathsf{G}^{+}{-}\mathsf{G}^{-})/2=-\pi \delta (E{-}\mathcal{H})$.
After some straightforward algebra we get:
\be\nonumber
\bm{G}^{kj} = 
i\frac{\hbar}{2\pi}\trc
\left[
\mathcal{F}^k
\mathsf{G}(E_F{-}i\Gamma/2)
\mathcal{F}^j
\im[\mathsf{G}(E_F)]
\right.
\\ -
\left.
\mathcal{F}^k
\im[\mathsf{G}(E_F)]
\mathcal{F}^j
\mathsf{G}(E_F{+}i\Gamma/2)
\right]
\ee

For the dot-wire geometry in the limit $L\rightarrow\infty$ 
we can treat the $i\Gamma$ as if it were the infinitesimal $i0$. 
Some more non-trivial steps allow us to reduce the trace 
operation to the boundary ($r=0$) of the scattering region (Fig.2), 
and then to express the result using the $S$ matrix.  
Disregarding insignificant interference term that 
has to do with having ``standing wave" the result is:
\be
\bm{G}^{3j} \ = \ \frac{e}{2\pi i}
\trc\left(P_{\tbox{A}}\frac{\partial S}{\partial X_j}
S^{\dag}\right)
\ee
This formula, which we derive here using ``quantum chaos" assumptions  
is the same as the BPT formula that has been derived for an open 
geometry. It is important to remember that the 
limit $L\rightarrow\infty$ is a non-adiabatic limit ($\Gamma\gg\Delta$). 
Still it is a ``DC limit". Therefore what we get here is 
``DC conductance" rather than ``adiabatic pumping". 
The latter term is unfortunately widely used in the existing literature.

\section{The prototype pumping problem}

What is the current which is created by translating 
a scatterer (``piston")? This is a ``pumping" question. 
Various versions of the assumed geometry 
are illustrated in Fig.4.  
Though it sounds simple this questions contains   
(without loss of generality) all the ingredients 
of a typical pumping problem. Below we address 
this question first within a classical framework, 
and then within quantum mechanics.

The simplest case is to translate a scatterer 
in 1D ring (Fig.4a). 
Assuming that there is no other scattering mechanism 
it is obvious that the steady state solution of 
the problem is:
\be
dQ = 1 \times \frac{e}{\pi}k_{\tbox{F}} \times dX
\ee
We assume here Fermi occupation, but otherwise 
this result is completely classical.
This result holds for any nonzero "size" of scatterer, 
though it is clear that in the case of a tiny scatterer  
it would take a much longer time to attain the steady state. 
Also note that there is no dissipation in this problem.
The steady state solution is an {\em exact} solution 
of the problem.

The picture completely changes if we translate 
a scatterer inside a chaotic ring (Fig.4b). 
In such case the problem 
does not possess a steady state solution. 
Still there is a quasi steady state solution.
This means that at any moment the state is 
quasi-ergodic: If we follow the evolution for 
some time we see that there is slow diffusion 
to other energy surfaces (we use here phase space 
language). This diffusion leads to dissipation 
as explained in \cite{frc} (and more Refs therein).
However, we are interested here mainly in the 
transport issue. As the scatterer pushes its way 
through the ergodizing distribution, it creates 
a current. Obviously the size of the scatterer 
{\em do matter} in this case. Using classical       
stochastic picture we can derive the following result:      
\be \label{e47}
dQ = 
\left[ \frac{g_T}{1{-}g_T}\right]
\left[ \frac{1{-}g_0}{g_0}\right]
\times
\frac{e}{\pi}k_{\tbox{F}} 
\times dX
\ee
where $g_0$ is the transmission or the relative size 
of the moving scatterer, while $g_T$ is the overall 
transmission of the ring.

What about the quantum mechanical analysis? 
We shall show that the same result is obtained 
{\em on the average}. This means that the 
classical expression still holds, but only 
in a statistical sense. This is in close analogy 
with the idea of ``universal conductance fluctuations". 
We shall discuss the effect of $\Gamma$ on the
distribution of~$\bm{G}$.

It should be noticed that our quantum chaos network 
model (Fig.4d)  essentially generalizes the two barrier model.  
Namely, one delta function is the ``scatterer" and 
the other delta functions is replaced by a complicated ``black box". 
Let us use the term ``leads" in order to refer to the two bonds 
that connect the ``black box" to the scatterer. 
Now we can ask what happens (given $\dot{X_1}$) if we take 
the length of the leads to be very very long. 
As discussed previously this is a non-adiabatic limit. 
We shall explain that in this limit we expect to 
get the same result as in the case of an open geometry.  
For the latter the expected result is \cite{AvronSnow}:  
\be \label{e48}
dQ = (1{-}g_0) \times \frac{e}{\pi}k_{\tbox{F}} \times dX
\ee
We shall explain how Eq.(\ref{e47}) reduces to Eq.(\ref{e48}).    
The latter is analogous to the Landauer formula $\bm{G}^{33}=(e^2/2\pi\hbar)g_0$.
The charge transport mechanism which is represented 
by Eq.(\ref{e48}) has a very simple heuristic explanation, 
which is reflected in the term ``snow plow dynamics" \cite{AvronSnow}.

\begin{figure}[b]
\epsfig{figure=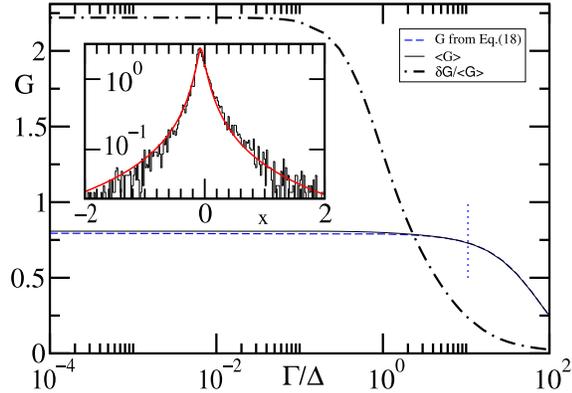,height=\hsize,angle=-90,clip}
\caption{   
The average conductance $\bm{G}^{31}$ for 
the network of Fig.4d. The average is 
taken over more than 20000 levels 
around $E_F$, while the calculation 
(for each Fermi level) was performed 
in an interval of 32000 levels. 
The transmission of the ``piston" 
is $g_{0} \approx 0.1$.  
The perpendicular dotted line indicates the border of 
the regime where the Kubo calculation is valid. 
We also plot the standard deviation,  
while the inset displays the distribution 
for $\Gamma=0.0001\Delta$. }
\end{figure}

\newpage
\section{Analysis of the network model}

One way to calculate $\bm{G}^{31}$ for the network model 
of Fig.4d is obviously to do it numerically using Eq.(\ref{e23}). 
For this purpose we find the eigenstates of the network, 
and in particular the wavefunctions $\psi^n=A_n\sin(k_nx+\varphi_n)$  
at (say) the right lead. Then we calculate the matrix elements 
\be
\mathcal{I}_{nm} &=& -i \frac{e\hbar}{2\mathsf{m}} 
\left(\psi^n\partial\psi^m - \partial\psi^n\psi^m\right)_{x{=}x_0} 
\\
\mathcal{F}_{nm} &=& 
-\lambda\frac{\hbar^2}{2\mathsf{m}}
\left(\psi^n\partial\psi^m + \partial\psi^n\psi^m
-\lambda \psi^n\psi^m \right)_{x=X_1+0}
\ee
and substitute into Eq.(\ref{e23}). The distribution 
that we get for $\bm{G}^{31}$, as well as the 
dependence of average and the variance on $\Gamma$ 
are presented in Fig.6. We see that $\Gamma$ 
reduces the fluctuations. If we are deep 
in the regime $\Delta \ll \Gamma \ll \Delta_b$ 
the variance becomes very small and consequently 
the average value becomes an actual estimate for $\bm{G}^{31}$.
This average value coincides with the ``classical" 
(stochastic) result Eq.(\ref{e47}) as expected on the 
basis of the derivation below.

In order to get an expression for $\bm{G}^{31}$ 
it is most convenient to use the FD expression Eq.(\ref{e37}).
For this purpose we have to calculate the 
cross correlation function of $\mathcal{I}$ and $\mathcal{F}^1$
which we denote simply as $C(\tau)$. If we describe 
the dynamics using a stochastic picture \cite{pmt} we get 
that $C(\tau)$ is a sum of delta spikes:
\be
C(\tau) \ = \ 
e\frac{v_{\tbox{F}}}{2L} 2\mathsf{m}v_{\tbox{F}}
\left[(1-g_0) \sum_{\pm} \pm \delta(\tau\pm\tau_1) \right]
+ {....} 
\ee
where $\tau_1 = (x_0-X_1) / v_{\tbox{F}}$ is the time 
to go from $X_1$ to $x_1$ with the Fermi velocity $v_{\tbox{F}}$, 
and the dots stand for more terms due to additional reflections.
If we integrate only over the short correlation 
then we get  
\be
\int_0^{\mbox{short}} C(\tau) d\tau \ \ = \ \ 
-e \frac{\mathsf{m}v_{\tbox{F}}^2}{L} 
\left[ 1-g_0 \right]
\ee
while if we include all the multiple reflections 
we get a geometric sum that leads to \cite{pmt}:
\be
\int_0^{\infty} C(\tau) d\tau \ \ = \ \ 
-e \frac{\mathsf{m}v_{\tbox{F}}^2}{L} 
\left[ \frac{1-g_0}{g_0}\right]
\left[ \frac{g_T}{1-g_T}\right]
\ee
This leads to the result that was already mentioned 
in the previous section:
\be \label{e54}
\bm{G}^{31} = -
\left[ \frac{1-g_0}{g_0}\right]
\left[ \frac{g_T}{1-g_T}\right]
\times \frac{e}{\pi}k_{\tbox{F}} 
\ee
We also observe that if the scattering in the outer 
region results in ``loss of memory",  then by Eq.(\ref{e38})
only the short correlation survives, and we get    
\be 
\bm{G}^{31} =  - (1-g_0) \times \frac{e}{\pi}k_{\tbox{F}}
\ee
Technically this is a special case of Eq.(\ref{e54}) 
with the substitution of the serial resistance 
$(1{-}g_T)/g_T = (1{-}g_0)/g_0 + (1{-}0.5)/0.5$.

The stochastic result can be derived also using 
a proper quantum mechanical calculation \cite{pmt}. 
The starting point is the following (exact) 
expression for the Green function: 
\be
\langle x | \mathsf{G}(E) | x_0 \rangle \ \ = \ \ 
- \frac{i}{\hbar v_F}\sum_p A_p \eexp{ik_E L_p}
\ee
The sum is over all the possible trajectories 
that connect $x_0$ and $x$. More details on 
this expression the the subsequent calculation 
can be found in Ref.\cite{pmt}. The final result 
for the {\em average} conductance coincides 
with the classical stochastic result.

\section{Summary}

Linear response theory is the major tool for study of 
driven systems. It allows to explore the crossover 
from the strictly adiabatic ``geometric magnetism" regime 
to the non-adiabatic regime. Hence it provides 
a unified framework for the theory of pumping. 

\begin{itemize}

\item[$\bullet$] 
``Quantum chaos" considerations in the derivation 
of the Kubo formula for the case of a closed isolated 
system are essential ($\Gamma\propto|\dot{X}|^{2/3}$).

\item[$\bullet$] 
We have distinguished between adiabatic, 
non-adiabatic and non-perturbative regimes, 
depending on what is $\Gamma$ compared with 
$\Delta$ and $\Delta_b$. 

\item[$\bullet$] 
In the strict adiabatic limit Kubo formula 
reduces to the familiar adiabatic transport 
expression (``geometric magnetism").

\item[$\bullet$]
A generalized Fluctuation-dissipation relation  
can be derived. In the zero temperature limit 
an implicit assumption in the derivation 
is having a generic bandprofile as implied 
by quantum chaos considerations.

\item[$\bullet$]
We also have derived an $S$ matrix expression 
for the generalized conductance of 
a dot-wire system, in the non-adiabatic 
limit $L\rightarrow\infty$. 
The result coincides with that of open 
system (BPT formula). 

\item[$\bullet$] 
The issue of ``quantized pumping" is analyzed 
by regarding the field which is created     
by ``Dirac chains". In the adiabatic regime 
$Q$ can be either smaller or larger than unity, 
while in the non-adiabatic regime $Q$ is less 
than unity in agreement with BPT.    

\item[$\bullet$]
We have analyzed pumping on networks 
using Green function expressions.
The average result can be expressed 
in terms of transmission probabilities. 
The analog of universal conductance fluctuations 
is found in the strict adiabatic regime. 
The conductance becomes well define (small dispersion) 
in the non-adiabatic regime.

\item[$\bullet$]
The average over the quantum mechanical result, 
which becomes the well defined conductance 
in the non-adiabatic regime, coincides with the 
result that had been obtained for the corresponding 
stochastic model.

\end{itemize}

\section{Acknowledgments}
I have the pleasure to thank T.~Kottos and 
H.~Schanz for fruitful collaboration,   
and Y.~Avishai, M.~B{\"u}ttiker, T.~Dittrich, 
M.~Moskalets and K.~Yakubo for discussions. 
This research was supported by 
the Israel Science Foundation (grant No.11/02),
and by a grant from the GIF, the German-Israeli Foundation 
for Scientific Research and Development.


\ \\ 

\texttt{http://www.bgu.ac.il/$\sim$dcohen/}

\end{document}